# Expert System Models in the Companies' Financial and Accounting Domain

Mates D., Iancu E., Bostan I., Grosu V.

**Abstract**— The present paper is based on studying, analyzing and implementing the expert systems in the financial and accounting domain of the companies, describing the use method of the informational systems that can be used in the multi-national companies, public interest institutions, and medium and small dimension economical entities, in order to optimize the managerial decisions and render efficient the financial-accounting functionality. The purpose of this paper is aimed to identifying the economical exigencies of the entities, based on the already used accounting instruments and the management software that could consent the control of the economical processes and patrimonial assets.

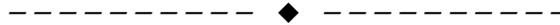

## 1 INTRODUCTION

Through this paper the authors have managed to show how, with the help of the expert system applied in the accounting domain economical and financial situations, offered by the classic systems can be examined, being able to alert, analyze and emit new decisions, that will either raise a question mark for the company's management for the decision that are mistaken or away from the optimum behavior, or confirm the fact that the existing decisions are the correct ones,

The expert applied systems recognize the fundamental role of the external information, making that the de-materialization and globalization of the informational processes are more and more accessible.

Within this context, the impact of the information is more and more extended, implying a greater number of investors or other economical and financial information users, interested differently, whose expectations have made an exponential raise in the pressure exerted over the entities, in terms of "chased results".

The intelligent systems used in accounting offer support to the management in optimizing the decisions regarding the direct and indirect cost management, with positive effects over the efficiency of the whole economical-financial activity (e.g.: the SYSCAT system).

Also with the help of the previously-described system consultancy can be offered, in contracting bank credits by the companies (e.g.: the SYBANK system), or realize the analysis of the economical- financial indices, so important for the accounting information users such as: investors, shareholders, managers, creditors, clients, suppliers, etc (e.g.: the DIASE system).

————————————————

- *Mates D. is with the West University of Timisoara, 16 Pezzdalozi Str., Timisoara, Romania.*
- *Iancu E. is with University of Suceava, 13 University Str., Suceava, Romania.*
- *Bostan I. is with the University of Suceava, 13 University Str., Suceava, Romania.*
- *Grosu V. is with the University of Suceava, 13 University Str., Suceava, Romania.*

The external information, therefore, is put in the center of a complex relation system between the entities, the complex and heterogenic palette of "stakeholders, characterized by various interests and objectives and the expert system models used varying with the management's necessities, having a decisive role in the equal development of the companies, dependent in the availability and optimum allocation of the financial resources necessary to generate competitive advantages and income producing.

The virtual world has changed the thinking way of the humans. That is why the companies must understand the high importance the expert systems have.

Shortly, if there is a modern and motivated management system, when there are the possibilities to make an operational expert system, the problem of making an expert system becomes similar to that of making an investment that must be lead by inspiration and desire that the expert system to be projected and implemented to be a practical, efficient, performing and useful system for the economical operator.

### RESEARCH METHODS

The expert systems developed with EXSYS Professional contain individual facts incorporated in decision taking knowledge pieces. These pieces are used in representing the knowledge with the aid of the production rule method. The EXSYS uses two types of facts (knowledge pieces): qualifiers and variables. Therefore, the knower must know that the EXSYS Professional operates with the following *base concepts*[1]:

- Rules- production rules
- Choices- purposes, decisional alternatives, recommendations

Qualifiers- or questions. The qualifiers are those knowledge pieces that allow the user to select one or more values from a list, predefined by the team made by the experts and knower. As a general rule, when a quali-fier is created, presenting the knowledge piece as a text that finishes with a verb, to which values are to be attached should be taking into consideration. Variables allow the user to introduce numerical or row values or that can be taken, through specialized interfaces, from the external programs/ applications or even Hypertext elements. For every newly-created variable it should be taken into consideration the fact that this one must be previously defined. It is very important to know that, because the text that helps in the description is taken and presented to the user in completing the standard message "please input a value for the variable"/ A variable can be used in any of the parts of a production rule. The name of a variable is written in brackets, and the attached message explains this name.

*The structure of the production rule and the work methods with the certainty factors*

The production rule in EXSYS Professional has 6 components:
  IF <premise>
  **THEN** <conclusion-1>
  ELSE <conclusion-2>
  **NOTE** <commentary>
  **REFERENCE** <commentary>
  NAME <name>

The last 4 components are optional. The IF part is made by combinations of qualifiers and associated values, The THEN part is made by combinations of alternatives/ purposes/ recommendations and numerical values, considered to be certainty factors. EXSYS offers 6 work methods with the certainty factors: *Yes/No, [0, 10], [-100, +100], Incr / Decr, Costum Formula* and Fuzzy. The left limits of each interval mean absolute uncertainty, and the right ones absolute certainty. The intermediate values indicate certainty factors that recommend action. Within a rule, if all the conditions of a premise are true, than the conclusion is true, that determining taking the rule in the attention of the inference engine for execution. The IF sentences such as sentences from other parts are English or Romanian phrases or even mathematical expressions. The THEN and ELSE contain alternatives for the possible solutions, that the EXSYS can select. The solutions are presented in a sentence followed by the certainty fact edited with the syntax **Confidence=<n>**, where <n> is a value from within the scale intervals presented previously, e.g.: 8/10, 5/10 in the case of the 0-10 scale. The EXSYS prefers first to infer the knowledge from other rules rather than solicit it from the user. This inference type is specific to the backwards control strategy. If the user addresses the question WHI?, during the consultation session there will be displayed, as an explication, the rules used in the inferential chain, When more explicative details are wanted, the „?" sign can be introduced. If a rule has been displayed, there is the possibility of asking the system from where he knows that the IF sentences are true, by typing the line corresponding to a condition. An explanation referring to one of the mathematical expression used can be requested, and therefore obtain the value of each involved variable. At the moment where the expert system reaches the conclusion/ solution of the problem, it displays a list (in descending order of the certainty attached factors) of the possible solution. Even notes or system evaluated variables' values can be displayed. After displaying the solutions, there is the possibility to change one or all the answers of the user in order to see the effect over the solutions.

**RESULTS AND DISCUSSIONS**

Within the economical domain, a lot of directions of possibilities to apply the future generations of computers are foreseen. Be it in the productivity domain, or in the distribution one, the operating systems through computer networks are successfully used for making orders to the suppliers, organizing the management in articles/ varieties, process the delivery orders from the customers, bill the delivered goods. The complete computerization of such types of activities means, obviously, lower costs, time save, quality services.

The expert systems, such as other categories of intelligent systems, manage to create certain myths regarding the way the solve problems and offer solutions, advice or recommendations. They have seducing qualities, in the way that they can provoke to some users an addiction towards them, being given that they are impressed by the way they work and give solutions, making them neglect the human interaction necessary with the managers from the superior level.

Clarke and Cooper, two British researchers, identify the future impact of the computerized technology over the accounting profession, with 4 types of changes:
  • Changes in the nature of the accounting labor, that meaning that a knowledge level will be created to encourage the involved personnel cooperating with the intelligent system designers
  • Changes in the complementary services, in the consultancy and training ways
  • Changes in the internal systems regarding projects planning, billing and human resources systems, because of the specialized software
  • Changes in the organizing structure, due to the raise of the personnel specialization and labor intensification at home in company interest.

About the accounting professionals it can be said that they are the most conservative and that they react slower to changes. But, it must still be taken into consideration that historically speaking the departments of accounting have nourished the most spectacular developments of computerized technology in enterprises.

It is said that the specialists within this domain must do the intellectual imagination test to benefit from the advantages of the technology and expert systems.

**The applicability area of the expert systems** in the financial accounting domain can be localized at the following activity segments:

- <u>Enterprise administration</u> - insures the fundamental financial balances and measurement of the enterprise's performances
- <u>For the employees</u> – analysis of the enterprise's stability, enterprise performance, payment systems
- <u>For the suppliers</u> - analysis of the financial situation and solvability
- <u>For the banks</u> – evaluating the credit granting risks
- Analysis of the solvability and credibility of potential customers

**The problems that can make the object of the expert systems** in financial accounting are:

- Choosing the most adequate account structures, that can answer to the decisional and informational requests;
- As informational requests, we include partners such as: banks, shareholders, customers;
- Taking data into the system, once with making the accounting analysis and proposing accounting recordings, according to the admitted principles;
- Taking data from the main documents through the expert systems will be accompanied by legality and consented operations opportunity check;
- Recognizing incomplete or mistaken data and treating these situations with specific procedures;
- Protection of the financial- accounting information with multiple access checking rules, at the data bases and the knowledge bases within the domain;
- Analysis and interpretation of the synthesis accounting documents: balance sheet, profit and loss account, annexes;
- Profitable leadership of the enterprise, using both accounting financial accounting information and management accounting information;
- Making complex works, such as: accounts consolidation or foreign relations accounting.

Within the management domain, three expert system classes can be described, at the level of the enterprises but also of the banks:

- Diagnostic expert systems, that imply making an expertise, based on rules;
- Prediction- planning expert systems, used to show the optimal plan variant;
- Control expert systems, for supporting decisions that must be taken extremely fast.

According to the Cambridge Dictionary, *to evaluate* means *to judge or calculate the quality, importance, amount or value of something*.

Taking into consideration this definition, we can consider *evaluating* as being the process of quantifying and expressing in money terms the assets, own capitals, debts, expenses, income, financial results, other assets and liabilities and all the events and transactions that have determined the modification of the financial position and enterprise's performances.

By *evaluation* are made the summation, grouping, centralization and generalization of the most diverse goods, events, activities and transactions, which allows *unitary expression* and comparison between various elements, expressed at first in quantity, that is in natural measurements or work.

In the accounting theory and practice, several *criteria* have taken shape, regarding the evaluation of flows or assets, liabilities, expenses and income stocks. These criteria are:
- Utility value
- Real value
- Market value
- Time

*a) The utility value* is represented by the price assumed to be accepted by an eventual buyer, varying with
- Use value
- Market price
- State or location of the element to be evaluated.

The utility value defines the value recognised by the seller and buyer within the direct transactions. The utility value must be regarded also from the losses point of vies, which an economical agent would support if not having a certain good absolutely necessary for its activity.

For a *liability element* (debts), the utility value is represented by the sums accepted to be paid in exchange for the created obligation, or the ones to be paid such as: taxes or other budgetary obligations.

*b) The just (real) value* is the sum at which an asset can be transacted or a debt can be discounted, willingly, within the parts that are in fully informed, within a transaction in which the price is objectively determined. It is based on form and content knowledge of the elements that are the object of the evaluation. In the case that *the acquisition or product cost* of an asset is not known, and there is no information regarding the prices or expenses necessary for its determination, or when such information can not be obtained without expense or unjustified delay, the acquisition or production cost will be represented by the *just (real) value* attributed to the asset.

*The acquisition cost* of a good, event, transaction consists of:
- Acquisition price
- Irrecoverable taxes
- Supplying transport expenses
- Other accessory expenses, necessary to put in usage state or management entering of the given good.

*The production cost* is made of:
- Acquisition cost of the raw materials and consumables;
- Other direct production costs
- Share part of the indirect production cost allocated rationally as being determined by the product fabrication.

*The general administration, financial and unpacking costs* are not included in the production costs, as an exception being the situations stipulated by the International Accounting Standards.

*c) The market value* of an asset represents the price that can be obtained on an active market. Its main *features* are:
-the market assets are relatively homogenous

-there are enough transacted assets, so that the possible buyers and sellers can be found at any time

-the prices are available for the public.

In reality, the market value is established trough the demand- offer report, on a completely free market, in which the offerers and buyers know in detail the parameters of the transaction.

*d) Time* is all about a certain calendar date, at which the evaluation is made. This can be : in the past, in the present or in the future. It is the consequence of the *activity continuation* process, according to which any activity, event or transaction comes from the past, goes through the present and produces effects in the future. According to the actual regulations, evaluating events and transactions at their entering in the patrimony is made based on the *historical cost* from the past that will generate future flows, events and transactions, materialized in assets and liabilities outcomes.

According to the International Accounting Standards, the *"assets"* represent the present resources controlled by the enterprise, as a result of past events, from which future benefits are expected.

The *liabilities* (debts) represent actual obligations of the enterprise, as a consequence of past events and by whose discount it is expected to result a resource outcome to incorporate economical benefits.

*Evaluation parameters in accounting*

The *evaluation parameters* are represented by independent variables, or own measurements of the goods, activities, events and transactions that allow the cash characterization of the assets, debts and own capitals.

According to the International Accounting Standards, the most representative parameters of evaluation are:
- the input value
- the net accounting value
- the current value
- the actualized value
- the realizable value
- the net realizable value
- the use value
- the recoverable value
- the residual value

*a) The input value* (cost) represents the cash expression of the assets and liabilities at the date of the entering in the unity, also known as the historical cost.

*The accounting value of the assets* represents the cash or cash equivalents that have been discounted to the suppliers at the time of the acquisition (minus VAT), or the production cost in the case of the ones resulted from own production

*The accounting value of the passives* represents the value of the equivalents obtained in exchange for the obligation or in certain conditions, the value that is expected to be paid in cash or cash equivalents to erase the debts according to the normal course of business.

The input values, as an evaluation parameter, can be expressed by:

-*the acquisition cost*, for the goods acquisitioned onerously (with payment)

-*the production cost*, for the goods resulted from own production

-*the value of intake,* for the goods brought as an intake to the social capital

-*utility value,* for the goods received as: subventions, free donation varying with the market value, stage and ampleness

-*nominal value,* for the sums probable to be received in exchange for the debts or to be paid in case of debts

-*correspondence value,* in the case of expense and income, established by association with the assets/ liabilities transformed in expense/ income.

Therefore, the *expenses* are evaluated either as a raise of the liability (in the case of expenses as debts) or a diminution of the asset (in the case of expenses as payments and consumptions).

*The income* is evaluated, as appropriate, as a raise of the asset, therefore as the value of the debt top the client, or as cash charged in case of selling, or the production cost in the case of income from own production of immobilizations and stocks.

In the case of asset outputs by selling, consuming etc, when there is not the possibility of identifying the input accounting value, the evaluation is made at:
- the average weight cost
- the price of the first input lot - in chronological order of lot exhaustion
- the price of the last input lot - in reverse chronological order of lot exhaustion

*b) The net accounting value* represents the sum at which an asset is recorded in the balance sheet, after discounting all the depreciations (amortizations, provisions, price differences etc.)

c) *The current value,* or the replacement cost is represented by the value equivalent that the enterprise accepts to pay, to get at the level of the present prices a good similar to the one shown as being the object of evaluation.

*d) The actualized value* represents a present estimation of the equivalents, future input cash flows generated by the asset elements, as a consequence of normally carrying on with an activity, pr the equivalent of the net future cash flows necessary to erase obligations, as a consequence of normal activity carrying on, in the case of the liabilities.

*e) The realizable value* is given by the cash equivalent, which can be obtained by present selling in normal conditions of the enterprise's assets.

*f) The net realizable value* is given by the sell price of an asset, less costs necessary for the sale.

*g) The use value* is represented by the actualized value of the future cash flows, estimated from common use of an asset and its cease at the end of the useful life period.

*h) The recoverable value* is the sum that the enterprise expects to recover from future use of an asset, including its residual value at the time of its alienation.

*i) The residual value* represents the net value that an enterprise estimates to obtain, by ceasing an asset at the end of its use period, after discounting the afferent cease costs.

The International and European Accounting Regulations use, as a base parameter in evaluation patrimonial assets and liabilities, the input value known as historical cost. Nevertheless, it is possible to combine with other parameters or use of alternatives based on the historical recoverable cost, or the concept of *maintaining the level of the physical or financial capital*. Therefore, the concept of *physical capital* must be adopted, if the main concern of the investors is the capacity of enterprise exploitation and the financial capital concept, in the case which the users of the financial situations are preoccupied with maintaining the nominal invested capital or its buying power.

## CONCLUSIONS

The virtual world has changed the human way of thinking. That is why the companies must understand the raised importance the expert systems have.

We can estimate that, globally, the next tendencies in the expert systems are manifested:

-making powerful KBSs (Knowledge Base Editors- source code or editor), perfectly adaptable to the given domain, based on which one can move forward to making expert systems

-making tandem systems that combine knowledge-based solutions with procedural solutions

-concept and domain notions' standardizing,

-coupling the expert systems with the data bases

Shortly, if there is a modern and motivated management system, when there are the possibilities to make an operational expert system, the problem of making an expert system becomes similar to that of making an investment that must be lead by inspiration and desire that the expert system to be projected and implemented to be a practical, efficient, performing and useful system for the economical operator.

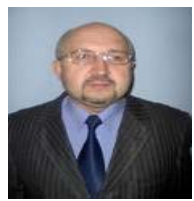
**Dorel Mates**, **Prof. PhD** -West University of Timisoara is a member in 6 academies and professional organisations, author and co-author of 7 books, over 50 scientifically papers, published abroad in specialty reviews or sustained and published within the international conferences and symposiums. He is a referee in the Science Committee of 2 foreign speciality journals and within the Science Committee of a foreign magazine.

**Ionel Bostan, Prof. PhD** - Stefan cel Mare University, authors 5 speciality books, over 10 papers published in 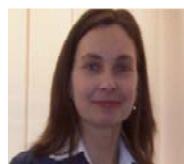 journals rated by ISI Thompson and 30 scientifically papers published in the country and abroad at the International Symposiums or Conferences. He is a member in 10 international professional organizations and scientifically referee in the editing committee of 2 journals rated by REPEC, Socionet, Research Gate etc, and within the scientifically committee of the WASET.

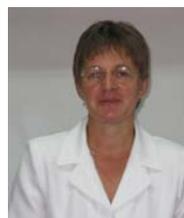**Veronica Grosu**, **Asistent PhD** - Stefan cel Mare University of Suceava is a co-author of 3 specialty books, over 10 papers published in journals rated by ISI Thompson or within the international symposiums and conferences. She is a member in 3 international professional organizations.

**Eugenia IANCU** Chair Lector of Informatics Desk, Economic Study and Public Administration Faculty, "Stefan cel Mare" University from Suceava. She is a doctorand at Technical Univesity from Timisoara. She's experienced in research contracts, she's part of the research team in 9 contracted projects, of which 4 are finalized and 5 are current.